\documentclass[a4paper,twocolumn]{article}
\usepackage{graphicx}

\def\be{\begin{equation}}
\def\ee{\end{equation}}
\def\bea{\begin{eqnarray}}

\def\eea{\end{eqnarray}}

\begin{document}

\title{Early Universe Quantum Processes in BEC Collapse Experiments}
\author{E. A. Calzetta$^1$ and B. L. Hu$^2$ \thanks{%
Emails: calzetta@df.uba.ar, hub@physics.umd.edu} \\
$^1${\small Departamento de Fisica, FCEyN Universidad de Buenos Aires Ciudad
Universitaria, 1428 Buenos Aires, Argentina}\\
$^2${\small Department of Physics, University of Maryland, College Park, MD
20742, USA}}
\date{{\small (March 11, 2005)}\\
\textit{\small - Invited Talk presented at the Peyresq Meetings of
Gravitation and Cosmology, 2003. To appear in Int. J. Theor.
Phys.}\\
 }
\maketitle

\paragraph{Main Theme}

We show that in the collapse of a Bose-Einstein condensate (BEC)
\footnote{For an excellent introduction to BEC theory, see
\cite{PS02}} certain processes involved and mechanisms at work
share a common origin with corresponding quantum field processes
in the early universe such as particle creation, structure
formation and spinodal instability. Phenomena associated with the
controlled BEC collapse observed in the experiment of Donley et al
\cite{JILA01b} (they call it `Bose-Nova', see also \cite{CVK03})
such as the appearance of bursts and jets can be explained as a
consequence of the squeezing and amplification of quantum
fluctuations above the condensate by the dynamics of the
condensate. Using the physical insight gained in depicting these
cosmological processes, our analysis of the changing amplitude
and particle contents of quantum excitations in these BEC
dynamics provides excellent quantitative fits with the
experimental data on the scaling behavior of the collapse time
and the amount of particles emitted in the jets. Because of the
coherence properties of BEC and the high degree of control and
measurement precision in atomic and optical systems, we see great
potential in the design of tabletop experiments for testing out
general ideas and specific (quantum field) processes in the early
universe, thus opening up the possibility for implementing
`laboratory cosmology'. \footnote{This essay has the same content
as v2 of \cite{CHbecPRL}, with a few references updated. For more
details, see \cite{CHbecPRA}.}

\paragraph{Theoretical Cosmology in relation to General Relativity, Particle
and Condensed Matter Physics}

For the last half a century the study of theoretical cosmology
and high energy astrophysics has relied largely on general
relativity and particle physics, while modern and contemporary
cosmological experiments are the fuse and the fuel of these
activities. We witness the inception of nuclear astrophysics in
the 50's, leading to the highly successful theories of neutron
stars, and particle astrophysics in the 60's exemplified by the
highly successful theory of nucleosynthesis which helped to
establish the standard model in cosmology. Establishment of
quantum field theory in curved spacetime \cite{BirDav} in the
70's laid the foundation for the study of quantum field processes
in strong gravity, such as cosmological particle creation in the
very early universe \cite{Schwinger} and Hawking radiation in
black holes \cite{Haw74} (Note these two processes contain very
different physics). This pushed the frontiers of theoretical
inquiries in leaps, back to the period after the Planck time. The
inflationary cosmology \cite{Guth} of the 80's also ushered in
ideas of particle physics and quantum field theory, such as the
decay of the false vacuum and the capability of vacuum energies
driving the universe into ultrafast expansion, with scenarios
radically different from the standard model.

This side of the story on the progress of modern cosmology with
the help of gravitation theory and particle physics is
well-known. What is perhaps lesser known or appreciated is the
importance of ideas and techniques from condense matter physics
and statistical mechanics in the study of cosmology of the early
universe. We have seen the relevance of statistical mechanics,
kinetic theory, stochastic processes and many-body dynamics in
classical astronomy and physical cosmology (see, e.g., \cite
{PeeblesPhysCos}). Here we want to emphasize the importance of
ideas from condensed matter physics in conjunction with quantum
field theory for treating early universe quantum processes, which
is believed to have played a fundamental role in determining how
spacetime and matter existing in different forms and states
interplay, transform and evolve. The importance of viewing
cosmology in the light of condensed matter physics, in terms of
taking the correct viewpoints to ask the right questions, and
approaches to understand the processes, has been called to our
attention a long time ago (see, e.g., \cite{HuHK}). There were
also proposals to study cosmological defect formation in helium
experiments and to view cosmology as a critical phenomenon \cite
{ZurekHe,SmoCos}. Similar efforts aim to identify analogs of full
cosmological models \cite{FF03}. A recent monograph is devoted to
the unity of forces at work in He$^3$ droplets \cite{Volovich}. It
should also be mentioned that the proposal of sonic black holes
\cite{UnrSonicBH,JacSonicBH} was perhaps the first analog model in
black hole physics which stimulated recent activities in finding
similar processes in fluids and condensed matter systems in the
so-called analog gravity program \cite{AnalogG}.

\paragraph{Laboratory Cosmology}

Here we propose using the Bose-Einstein Condensate (BEC) and its
dynamics as another useful venue to `observe' and probe into some
fundamental cosmological processes in the early universe.
Specifically, we analyze the experiment performed by Donley et
al. \cite{JILA01b} on the controlled collapse of a BEC and
identify the processes and mechanisms at work which are
responsible for vacuum particle creation \cite{Schwinger},
structure formation \cite{strfor} and spinodal instability
(quenching) in phase transition \cite{SD} in the early universe.
The collapsing BEC is the time-reverse scenario of an expanding
universe and the condensate plays a similar role as the vacuum in
quantum field theory in curved spacetime. One can understand the
production of atoms in the form of jets and bursts as the result
of parametric amplification of vacuum fluctuations by the
condensate dynamics. This is the same mechanism as cosmological
particle creation from the vacuum, which is believed to be
copious near the Planck time and during preheating after
inflation \cite{preheat}. Some basic ideas common to cosmological
theories like ``modes freeze when they grow outside of the
horizon'' can be used to explain the special behavior of jets and
bursts ejected from the collapsing BEC. Finally the waiting time
before a BEC starts to collapse obeys a scaling rule which can be
derived from simple principles of spinodal instability in
critical phenomena. These examples clearly indicate the great
potential of a new field of research which we may call
``laboratory cosmology'', with tabletop experiments designed to
test the workings of specific physical mechanisms in specific
cosmological processes.

\paragraph{BEC Collapse Experiments}

In the experiment described by Donley et al. \cite{JILA01b}, a
Bose-Einstein condensate (BEC) in a cold ($3$nK) gas of Rubidium
atoms is rendered unstable by a sudden inversion of the sign of
the interaction between atoms. This is done by altering the
binding energy at Feshbach resonance with an external magnetic
field. After a waiting time $t_{collapse},$ the condensate
implodes, and a fraction of the condensate atoms are seen to
oscillate within the magnetic trap which contains the gas. These
atoms are said to belong to a `burst'. After a time $\tau
_{evolve}$ the interaction is suddenly turned off. For a certain
range of values of $\tau _{evolve},$ new emissions of atoms from
the condensate are observed. They are called `jets'. Jets are
distinct from bursts: they are colder, weaker,
and have a characteristic disk-like shape. \footnote{%
We call attention to the distinction between the 'Bose-Nova' \cite{JILA01b}
experiment studied here and other BEC collapse experiments \cite
{JILA98,JILA02a}. At magnetic fields around $160$G, where the effective
scattering length is of the order of $500a_{0}$ (and positive)($a_{0}=0.529$%
\ $10^{-10}$m\ \ \ is the Bohr radius) it is possible to observe
oscillations between the usual atomic condensate and the molecular state
\cite{JILA02a} with a frequency of oscillations of hundreds of KHz \cite
{KGB02}. By contrast, in the 'Bose-Nova' experiment \cite{JILA01b} typical
fields were around $167$G, the scattering length was only tens of Bohr radii
(and negative) and the frequency of atom - molecule oscillations may be
estimated as well over ten MHz \cite{JILA03}. While coherent resonance
between the atoms and the molecules is expected to exist for all of these
experiments, and has been shown to play an important role in the outcomes of
some \cite{JILA03}, we deem it unlikely that it plays a dominant role in
this experiment other than renormalizing the scattering length (For details,
see \cite{CHbecPRA}). Indeed no oscillations are reported in the original
experimental paper. Instead, as this note shows, the primary mechanism for
the Bose Nova phenomena is the parametric amplification of quantum
fluctuations by the condensate dynamics, resulting in bursts and jets as
particle production from (the squeezing of) the vacuum. Recent numerical
simulations \cite{Savage} and rigorous theoretical investigations \cite{DS03}
indicating the inadequacy of mean field theory seem to corroborate this view.%
}

\paragraph{The Model}

The model is based on the Hamiltonian operator for $N$ interacting atoms
with mass $M$ in a trap potential $V\left( \mathbf{r}\right) =(\omega
_{z}^{2}z^{2}+\omega _{\rho }^{2}\rho ^{2})/2$, with radial $\rho $ and
longitudinal $z$ coordinates measured in units \footnote{%
We use a sign convention such that the effective coupling constant is
positive for an attractive interaction, and a system of units where the
length $a_{ho}$, time $t_{ho}$ and energy scale $E_{ho}=\hbar \omega
=M\omega ^{2}a_{ho}^{2}$ are defined with reference to the average frequency
$\omega $. We work with units such that these three scales take the value $1$%
.} of $a_{ho,}$ where $a_{ho}$ is a characteristic length of the trap, with
associated (dimensionless) frequencies $\omega _{z}=\omega _{axial}/\omega
\sim 1/2$ and $\omega _{\rho }=\omega _{radial}/\omega \sim \sqrt{2}.$ The
interaction is assumed to be short ranged. We introduce a dimensionless
field operator $\mathbf{\Psi }\left( r\right) \equiv a_{ho}^{-3/2}\Psi
\left( x\right) $, and a dimensionless coupling constant $u=\left( \hbar
\omega a_{ho}^{3}\right) ^{-1}U=4\pi \left( a/a_{ho}\right) $.

$\Psi $ obeys the equation of motion $\dot{\Psi}=i\left[ \hat{H},\Psi
\right] $ and satisfies the equal time commutation relations $\left[ \Psi
\left( t,\mathbf{r}\right) ,\Psi ^{\dagger }\left( t,\mathbf{r^{\prime }}%
\right) \right] =\delta ^{\left( 3\right) }\left( \mathbf{r}-\mathbf{%
r^{\prime }}\right) .$ We decompose the Heisenberg operator $\Psi =\Phi (%
\mathbf{r},t)+\psi (\mathbf{r},t)$ into a c-number condensate amplitude $%
\Phi $ and a q-number noncondensate amplitude $\psi $, consisting of the
fluctuations or excitations.

One crucial point of our analysis is that we shall focus on the evolution of
the fluctuations for a given evolution of the condensate (as extracted from
the experiments); in field theory terms, we shall work within the test field
approximation. It is fair to say that a full theoretical account of the Bose
- Nova experiment, describing the evolution of both condensate and
fluctuations, does not exist. All we can say with any certainty is that we
face here a strong back reaction regime, beyond the Hartree - Fock -
Bogoliubov approximation \cite{WHS04}.

In contradistinction, the dynamics of the fluctuations alone may be
described, at least in the early stages of the experiment, by a simple
Bogoliubov approximation \cite{AND04}. We obtain the equation of motion for
the fluctuation field by subtracting from the full Heisenberg equation the
Gross - Pitaievsky equation (GPE) for $\Phi $. We next parametrize the wave
functions as $\Phi =\Phi _{0}e^{-i\Theta },$ $\psi =\psi _{0}e^{-i\Theta }$,
where $\Phi _{0}$ and $\Theta $ are real. During the early stages of
evolution, we may regard the condensate density as time independent, and the
condensate phase as homogeneous, $\Phi _{0}=\Phi _{0}\left( r\right) ,$ $%
\Theta =\Theta \left( t\right) $. We may then write the equation for the
fluctuation field
\begin{equation}
\left[ i\frac{\partial }{\partial t}-H+E_{0}\right] \psi _{0}+u\Phi
_{0}^{2}\left( \psi _{0}+\psi _{0}^{\dagger }\right) =0  \label{dyn}
\end{equation}
where $E_{0}=\frac{1}{2}\left( \omega _{z}+2\omega _{\rho }\right) $. As it
is well known, this approximation is both ``gapless'' and ``conserving''
\cite{gapless}.

To solve equation (\ref{dyn})\footnote{%
The squeezing of quantum unstable modes and its back reactions on the
condensate has been considered before, e.g., as a damping mechanism for
coherent condensate oscillations \cite{KM01}, and applied to the collapse of
a homogeneous condensate  in \cite{Y02}.} we decompose $\psi _{0}$ into a
self-adjoint and an anti-adjoint part $\psi _{0}=\xi +i\eta $, each part
satisfying an equation
\begin{equation}
\frac{\partial \xi }{\partial t}=\left[ H-E_{0}\right] \eta
\label{treintaydos}
\end{equation}

\begin{equation}
\frac{\partial \eta }{\partial t}+\left[ H-E_0-2u\Phi _0^2\right] \xi =0.
\label{treintaytres}
\end{equation}

Since the trap Hamiltonian is time - independent, we have

\begin{equation}
\frac{\partial ^{2}\xi }{\partial t^{2}}+\left[ H-E_{0}\right] H_{eff}\xi =0.
\label{treintaycuatro}
\end{equation}
Here $H_{eff}=H-E_{0}-2u\Phi _{0}^{2}$. To have an unstable condensate it is
necessary that at least one of the eigenvalues of $H_{eff}$ is negative; the
boundary of stability occurs when the lowest eigenvalue is exactly zero.

One further consideration is that we are interested in the part of the
fluctuation field which remains orthogonal to the condensate. In the full
theory, the condensate is the eigenfunction of the one - body density matrix
with the largest (macroscopic) eigenvalue, and the non-condensate is built
out of the other eigenstates \cite{PO56}. Since the one-body density matrix
is Hermitian, they must be orthogonal. The ground state of $H_{eff}$ is
certainly not orthogonal to the condensate, since neither have nodes.
Observe that within our approximations, the noncondensate wave function is
equivalent, up to a normalization, to the phonon operator in the particle-
conserving formalism \cite{conserving}.

If we adopt the values $\omega _{z}=1/2,$ $\omega _{\rho }=\sqrt{2},$
relevant to the JILA experiment, then instability occurs when $\kappa
=N_{0}a_{crit}/a_{ho}=0.51$. This result compares remarkably well with the
experimental value $\kappa =0.55$ \cite{JILA01b,JILA03}, as well as with the
theoretical estimate presented in Ref. \cite{GTT01}. This agreement may be
seen as natural, as the equations we postulate for the fluctuations may be
obtained from the linearization of the GPE. In both calculations, the
geometry of the trap plays a fundamental role.

\paragraph{Scaling of $t_{collapse}$ and Critical Dynamics}

As we have already noted, even for condensate densities above the stability
limit, no particles are lost from the condensate during a waiting time $%
t_{collapse}.$ Experimentally, $t_{collapse}$ is seen to get very large when
the threshold of stability is approached from above, in a way which closely
resembles the \textit{critical slowing down} near the transition point
characteristic of critical dynamics. In our problem, the quantity which
plays the role of relaxation time is the characteristic time $\varepsilon
^{-1}$ of exponential growth for the first unstable mode. This quantity
diverges at the stability threshold, which in our analogy corresponds to the
critical point. By dimensional analysis, we are led to the estimate $%
t_{collapse}\sim \varepsilon ^{-1}.$ Close to the critical point, we find

\begin{equation}
t_{collapse}=t_{crit}\left( \frac{a}{a_{cr}}-1\right) ^{-1/2}
\label{scaling}
\end{equation}

The power law Eq. (\ref{scaling}) describes with great accuracy the way $%
t_{collapse}$ scales with the scattering length; the best fit to the
experimental data is obtained for $t_{crit}\sim 5$ms.

\begin{figure}[h]
\includegraphics[height=4cm]{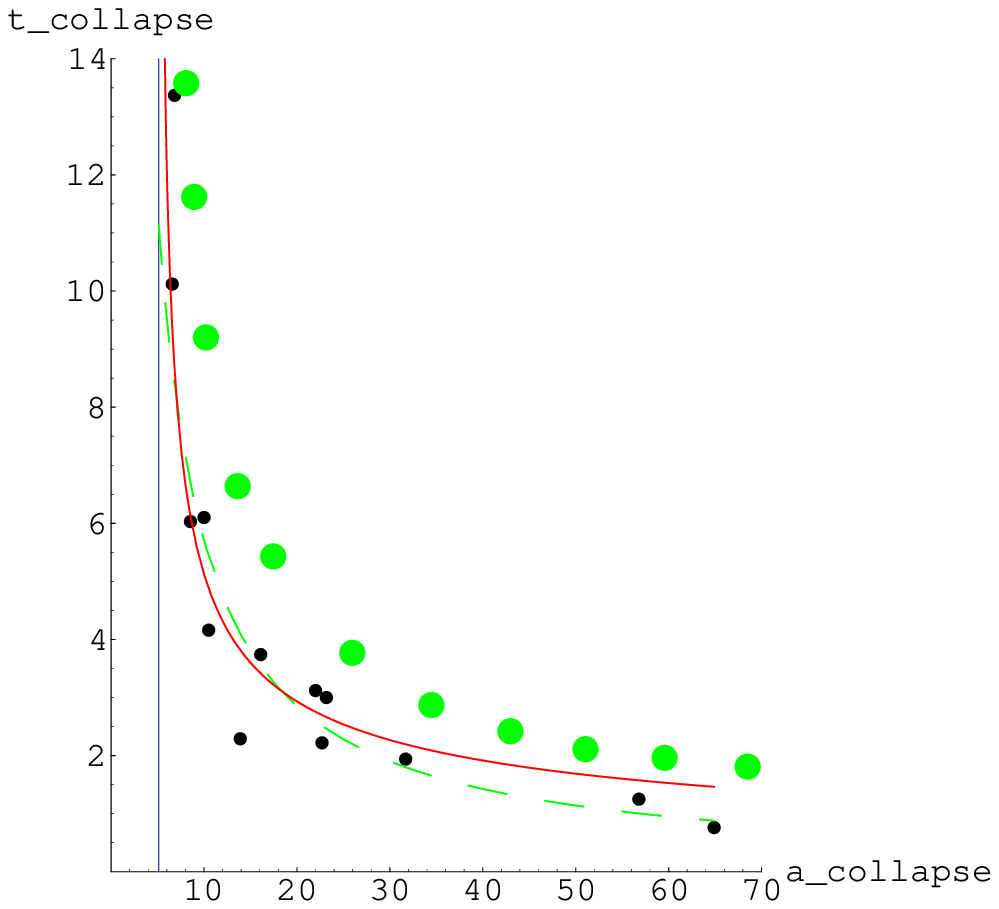}
\caption{}
\end{figure}

In Fig. 1 we plot the scaling law (\ref{scaling}) (full line) derived here
and compare it with the experimental data for $N_{0}=6000$ as reported in
Refs. \cite{JILA01b} (small black points), the $t_{NL}\sim \left(
uN_{0}\right) ^{-1}$ prediction (suitably scaled) as given in \cite
{Y02,TBJ00} (dashed line) and the results of numerical simulations reported
in \cite{SU03} (large grey dots). While all three theoretical predictions
may be considered satisfactory, the $t_{NL}\sim \left( uN_{0}\right) ^{-1}$
behavior fails to describe the divergence of $t_{collapse}$ as the critical
point is approached, and the results of numerical simulations reported in
\cite{SU03} based on an improved Gross-Pitaevskii equation tend to be
systematically above the experimental results, which may be a further
indication of the quantum origin of this phenomenon \cite{BSB02}.

We wish to stress that our argument predicts the scaling exponent, but not
the prefactor; even this apparently simple aspect of the Bosenova
phenomenology is surprisingly resilient to theoretical explanation \cite
{WHS04}. The same scaling law is found from a different perspective in \cite
{MDB03}.

\paragraph{Bursts and Jets as Amplified Quantum Fluctuations}

We now consider the evolution of quantum fluctuations, treated as a test
field riding on the collapsing condensate whose dynamics is extracted from
experiment. The initial state is defined by the condition that $u=0$ for $%
t<0;$ we shall take it to be the particle vacuum $\left| 0\right\rangle$,
defined by $\psi _0\left( x,0\right) \left| 0\right\rangle =0$ everywhere.

One can introduce a mode decomposition of the $\xi $ operator based on the
eigenfunctions of $\left[ H-E_{0}\right] H_{eff}$. For short wavelengths $%
\lambda $, since $H\sim \lambda ^{-2}>>2u\Phi _{0}^{2}$, we expect these
eigenfunctions will approach the trap eigenmodes. The fact that particles in
bursts are seen to oscillate with the trap frequencies \cite{JILA01b} also
suggests that their dynamics is determined by the trap Hamiltonian. Based on
these observations we can assume a homogeneous condensate $2u\Phi
_{0}^{2}\sim \kappa ^{-1}a\omega _{z}N_{0}\left( t\right) $, where $%
N_{0}\left( t\right) $ is the instantaneous total number of particles in the
condensate. In practice, $\kappa ^{-1}$ is a measure of the overlap between
the condensate and the excitation modes. Therefore, the approximation may be
improved by adjusting $\kappa $ according to the range of modes where it
will be applied.

Let $\bar N_0$ be the initial number of particles in the condensate, and $%
a_{cr}=\kappa /\bar N_0$ the corresponding critical scattering
length. Trap eigenfunctions $\psi _{\vec n}\left( r\right) $ are
labeled by a string of quantum numbers $\vec n=\left(
n_z,n_x,n_y\right) .$ The eigenvalues of the trap Hamiltonian are
(with the zero energy already subtracted) $E_{\vec n}=\omega
_zn_z+\omega _\rho \left( n_x+n_y\right) $. There are two kinds of
modes, stable (oscillatory, or thawed) modes if $E_{\vec
n}>\left( \frac a{a_{cr}}\right) \omega _z,$ and unstable
(growing, or frozen) modes if not. In the former case we find
that, although we assume vacuum initial conditions, these modes
do not remain empty. Up to $t_{collapse}$, when the number of
particles in the condensate is constant, the density

\begin{equation}
{\tilde n}\left( r,t\right) =\frac 18\left( \frac a{a_{cr}}\right) ^2\omega
_z^2\sum_{\vec n}\psi _{\vec n}^2\left( r\right) \frac{\sin ^2\omega _{\vec
n}t}{\omega _{\vec n}^2}\   \label{thawed}
\end{equation}
(where $\omega _{\vec{n}}=\sqrt{E_{\vec{n}}\left[ E_{\vec{n}}-\left( \frac{a%
}{a_{cr}}\right) \omega _{z}\right] }$) has a constant term and an
oscillatory term. This oscillatory term is responsible for the appearance of
`\textbf{bursts}' of particles oscillating within the trap observed in the
Bose-Nova experiment \cite{JILA01b}. In the WKB limit it describes a swarm
of particles moving along classical trajectories in the trap potential.

In the opposite case $E_{\vec n}\le \left( \frac a{a_{cr}}\right) \omega _z,$
the formulae for the density is obtained by the replacement of $\omega
_{\vec n}$ in (\ref{thawed}) by $i\sigma _{\vec n}$, thus $\omega _{\vec
n}^{-1}\sin \omega _{\vec n}t\rightarrow \sigma _{\vec n}^{-1}\sinh \sigma
_{\vec n}t.$ Physically their difference is immense. In the first place, the
density is growing exponentially, but unlike the previous case, there is no
oscillatory component, and these particles do not oscillate in the trap, in
the sense described above. These modes come alive at $\tau _{evolve}$ (as
the scattering length is set to zero), whence they become ordinary trap
modes which oscillate in the trap in the same way as the the burst modes .
To the observer, they appear as a new ejection of particles from the core of
the condensate, which makes up the so-called `\textbf{jets}'. The sudden
activation of a frozen mode (we are borrowing the language and concept of
cosmological structure formation) by turning off the particle - particle
interaction may be described as a ``thaw''.

Observe that in this picture several conspicuous features of jets become
obvious. Jets may only appear if the turn - off time $\tau _{evolve}$ is
earlier than the formation time of the remnant. Once the condensate becomes
stable again, there are no more frozen modes to thaw. On the other hand,
jets will appear (as observed) for $\tau _{evolve}<t_{collapse}$, when the
condensate has not yet shed any particles. Also jets must be less energetic
than bursts, since they are composed of lower modes.

Beyond $t_{collapse}$ the number of particles in the condensate, and
therefore the instantaneous frequency of the excited modes, becomes time
dependent. If we confine ourselves to the early stages of collapse we may
assume nevertheless that the condensate remains homogeneous. Shifting the
origin of time to $t_{collapse}$ for simplicity, we write $N_0\left(
t\right) =\bar N_0\mathrm{exp}\left( -t/\tau \right) $ (see Fig. 2).

\begin{figure}[h]
\includegraphics[height=3cm]{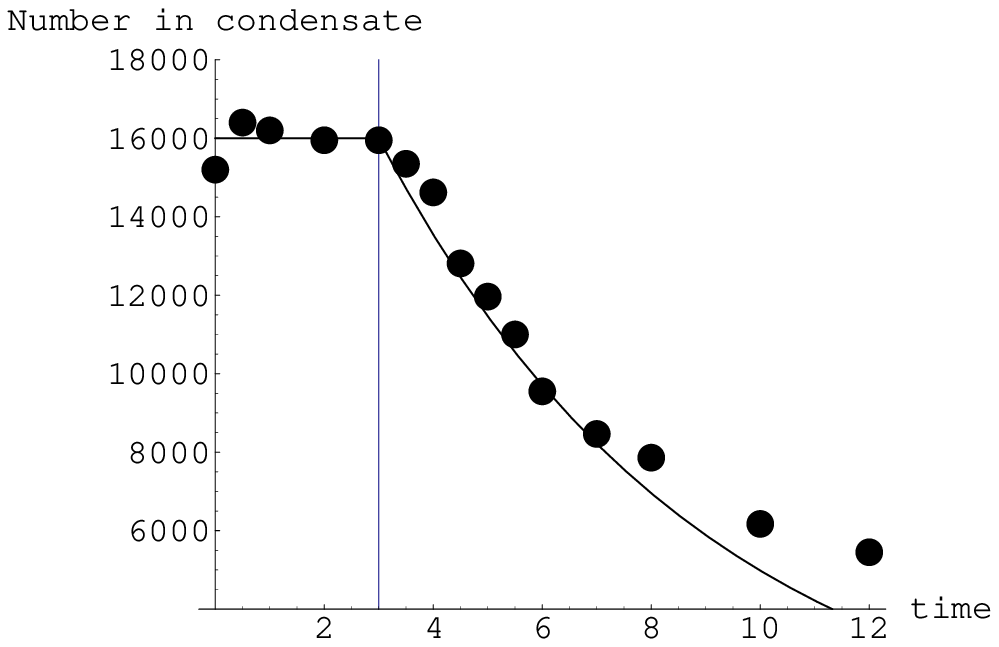}
\caption{{}}
\end{figure}

After expanding in trap eigenmodes we find the two kinds of behavior
described above. If $E_{\vec n}>\left( \frac{a\omega _z}{\bar a}\right) ,$
the mode is always oscillatory. If $E_{_{\vec n}}<\left( \frac{a\omega _z}{%
\bar a}\right) ,$ the mode is frozen at $t_{collapse},$ but thaws when $%
\mathrm{exp}\left( -t/\tau \right) \sim E_{\vec n}\bar a/a\omega _z$. During
the frozen period, the modes are amplified, but they only contribute to
bursts after thawing. If the evolution is interrupted while still frozen,
they appear as a jet. We therefore conclude that the number of particles $%
N_{jet}$ in a jet at time $\tau _{evolve}$ is essentially the total number
of particles in all frozen modes at that time. This is plotted in Fig 3, for
$\bar{N}_{0}=16,000,$ $\omega _{radial}=110$ Hz, $\omega _{axial}=42.7$ Hz, $%
a=36a_{0},$ and $\kappa =0.46$ , and compared to the corresponding results
as reported in \cite{JILA01b}.

\begin{figure}[h]
\includegraphics[height=4cm]{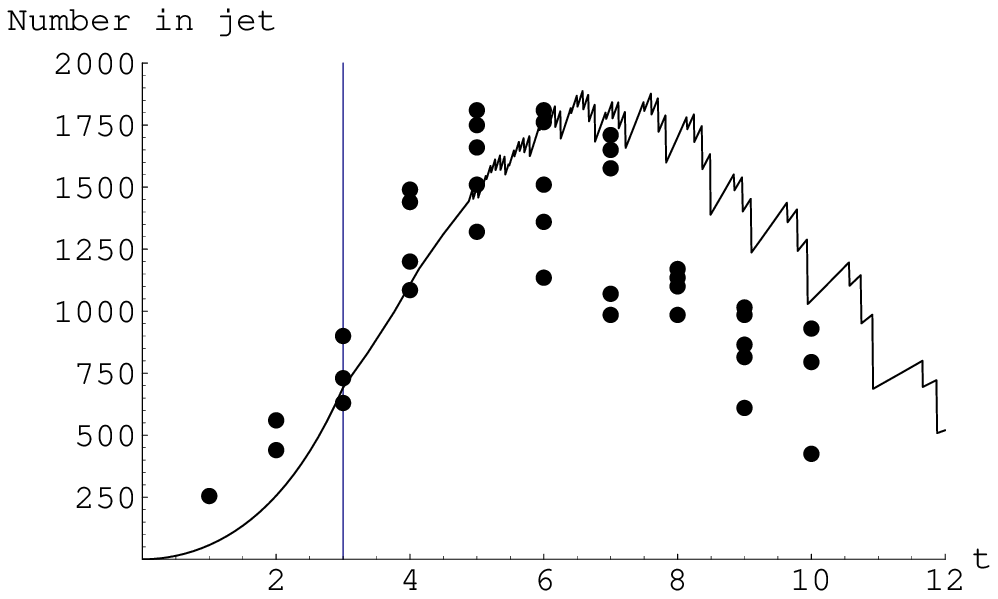}
\caption{}
\end{figure}

We see that the agreement is excellent at early times (up to about $6$ms).
For later times, this model overestimates the jet population. This is due to
the fact that, by not considering the shrinking of the condensate, we are
overestimating the overlap between the condensate and the fluctuations, thus
delaying the thaw. It nevertheless reproduces the overall slope of particle
number with $\tau _{evolve},$. It should also be remembered that we are
computing the expected number of particles, but in the highly squeezed state
which results from the frozen period, the fluctuations in particle number
are comparable to the mean number itself.

It is interesting to observe that it is possible to reproduce the jets
within a theory where the GPE\ equation is generalized to include an
imaginary three - body recombination loss term \cite{SS02}. We do not regard
this as necessarily a different explanation, but rather as a different way
of handling the divide between condensate and non-condensate (the analogous
cosmological problem would be whether to consider a spin two fluctuation
generated during Inflation as a test field on the cosmological background,
or as part of the geometry). Still agreement with the observed jets is
obtained only for certain ranges of parameters, and it is unclear whether
any single parameter set gives a satisfactory simultaneous account of all
aspects of the experiment \cite{WHS04}.

In this talk, we have presented a new viewpoint towards
understanding the salient features in the physics of controlled
collapse of a Bose-Einstein condensate described in the
experiment of \cite{JILA01b}, i.e, in terms of quantum vacuum
fluctuations parametrically amplified by the condensate
dynamics.  Even under a number of simplifying assumptions, our
approach yields results in excellent agreement with experiment,
particularly in the scaling of the waiting time $t_{collapse}$
and the number of particles in a jet. A background field
separation is assumed here (even though the measured condensate
dynamics contains the backreaction of noncondensates) so that one
can treat these as test-field processes. A theoretical treatment
of the fully self-consistent dynamics of both the condensate and
its quantum fluctuations during collapse beyond the conventional
Hartree - Fock - Bogolubov theory  remains a worthy challenge
(see, e.g., \cite{REY04} and references therein).

If our explanation of the salient features of this experiment is correct one
can think of using this process to create coherent atoms in highly squeezed
states \cite{squeeze}. This is because the underlying mechanism of
parametric amplification produces particles from vacuum fluctuations in
squeezed states \cite{squeezecos}.

Our way of thinking here is influenced by insights from the
quantum field theory of particle creation and structure formation
in cosmological spacetimes as well as theories of spinodal
instability in phase transitions. One can conceivably design
experiments with BEC dynamics to test out certain basic
mechanisms and specific features of quantum processes in the early
universe, thus opening a new venue for performing `laboratory
cosmology'.

\paragraph{Acknowledgement}

We thank Bill Phillips and Keith Burnett for kindling our
interest in BEC dynamics, Elizabeth Donley and Jake Roberts for
clarification of their data in the experiments described here and
S. Kokkelmans for communicating key unpublished data (at the time
the letter \cite{CHbecPRL} on which this essay is based was
written). We thank our colleagues Ted Jacobson, Stefano Liberati
at Maryland and Charles Clark and his group members at NIST
Gaithersburg during our seminars for their questions and
comments. EC acknowledges discussions with Eric Bolda. This
research is supported in part by NSF grant PHY03-00710 and by
CONICET, UBA, Fundacion Antorchas and ANPCyT under grant PICT99
03-05229.

\end{document}